\documentstyle[12pt,epsfig]{article}

\newcommand{\be}{\begin{equation}}
\newcommand{\bea}{\begin{eqnarray}}
\newcommand{\eea}{\end{eqnarray}}
\newcommand{\ba}{\begin{array}}
\newcommand{\ea}{\end{array}}
\newcommand{\ee}{\end{equation}}

\newcommand{\cN}{{\cal N}}
\newcommand{\cR}{{\cal R}}

\newcommand{\RR}{\mbox{\boldmath$R$}}

\expandafter\ifx\csname mathbbm\endcsname\relax

\else

\fi
\def\appendix{{\newpage\section*{Appendix}}\let\appendix\section%
        {\setcounter{section}{0}
        \gdef\thesection{\Alph{section}}}\section}

\begin{document}
\begin{titlepage}
\hfill
\vbox{
    \halign{#\hfil         \cr
           CERN-TH/2000-351 \cr
           hep-th/0011288  \cr
           } 
      }  
\vspace*{30mm}
\begin{center}
{\Large {\bf  Supergravity and D-branes Wrapping Supersymmetric
3-Cycles}\\} 

\vspace*{15mm}
\vspace*{1mm}
{Harald Nieder  and Yaron Oz}\\

\vspace*{1cm} 

{\it Theory Division, CERN \\
CH-1211, Geneva, 23, Switzerland}\\

\vspace*{1cm}
\end{center}

\begin{abstract}
We construct dual supergravity descriptions of D3-branes wrapping associative 3-cycles $L$.
We analyse the conditions
for having 
five-dimensional background solutions of the form $AdS_2 \times L$ 
and show that they require
$L$ to be of constant negative
curvature type. This provides $AdS_2$ background solutions
when $L$  is the hyperbolic space $H^3$ or
its quotients by subgroups of its isometry group.
We construct a regular numerical solution interpolating between $AdS_5$ in the UV
and  $AdS_2 \times H^3$ in the IR.
The IR fixed point exists at the ``intersection'' of the Coulomb and Higgs branches.
We analyse the singular supergravity solutions which correspond
to moving into the Higgs and the Coulomb branches.
For negative  constant curvature spaces the singularity is of a ``good'' type
in the  Higgs branch and of a ``bad'' type
in the Coulomb  branch.
For positive constant curvature spaces 
such as $S^3$ the singularity is of a ``bad'' type
in both the  Higgs and
the Coulomb  branches.
We discuss the meaning of these results.
\end{abstract}
\vskip 3.5cm

November 2000
\end{titlepage}

\newpage

\section{ Introduction}

Supersymmetric 3-cycles 
are defined such 
that D-branes wrapping them
are supersymmetric configurations \cite{BBS,BSV,OOY,BB}.
We will discuss two types of supersymmetric 3-cycles: associative 3-cycles in $G_2$
holonomy seven-dimensional manifolds and special Lagrangian 3-cycles 
in Calabi-Yau 3-folds 
\footnote{For a discussion and
review of special Lagrangian 3-cycles and their role in ``brane worlds'' and
curved geometries see \cite{shamit}.}.
The low-energy field theory on the worldvolume
of Dp-branes wrapping associative (special Lagrangian) 3-cycles is a $(p-2)$-dimensional
gauge theory with two (four) supercharges.
The Dp-brane worldvolume theory 
is a twisted field theory \cite{BSV}. 
The holonomy group of a  3-cycle $L$ is $\subseteq SO(3)$ and its
 normal bundle is non-trivial.
The  scalars parametrizing the embedding of a Dp-brane
wrapping $L$ in
a the ambient space $M$ 
 are sections of the normal bundle of $L$ in $M$ and are not
ordinary scalars. 
The 
twisting means choosing the holonomy group of the tangent bundle 
to be equal to the structure (sub)group of the normal bundle.
This  gives covariantly constant supercharges
which 
are ``twisted'' and become scalars rather than spinors.

We will consider Type IIB string theory compactified on 
$M$ and
$N$ D3-branes wrapping the 3-cycle $L$ in the limit
$l_s \rightarrow 0$ where we keep the volumes of the cycle $L$ and the 
manifold $M$ fixed \cite{malda}.
While at high energies the worldvolume theory is four-dimensional,
at low energies compared to the inverse size of $L$ the theory is $(0+1)$-dimensional.
We will be 
interested in the infrared (large time)
behaviour of the quantum mechanics.

Our aim is to construct  dual supergravity descriptions
of the 
worldvolume theories of the
D3-branes wrapped around supersymmetric 3-cycles.
Supergravity descriptions of branes wrapping supersymmetric cycles have been constructed
in various cases such as for Dp-branes and M5-branes wrapping 
Riemann surfaces in Calabi-Yau spaces \cite{AO,CV,FS,malda,malda2,KS}
and M5-branes wrapping associative cycles in $G_2$-holonomy manifolds \cite{g2}.
Here the twist induces the boundary conditions 
on the gauge field coming from the metric upon compactification \cite{malda}.
We will use a five-dimensional truncation of the full type IIB supergravity. 
Since to implement the 
twist we need an $SO(3)$ gauge potential, 
one suitable supergravity theory in five dimensions 
is the $\cN=4$ $SU(2) \times U(1)$ gauged supergravity described in \cite{romans}
 which was shown to be a consistent truncation of type IIB supergravity in \cite{pope}. 
We will see
that this gauged supergravity can realize the twist required
for D3-branes wrapping an associative 3-cycle.

To analyse
what types of 3-cycles we should consider, we note
the eight geometric types in the classification of closed 3-manifolds
introduced by 
Thurston  \cite{t}.
The most relevant for us are the first 
three types in the classification,
based on the three constant curvature spaces, the 3-sphere $S^3$
which has a positive scalar curvature $R >0$
and isometry group $G=SO(4)$, the Euclidean space  $E^3$ with  $R = 0$ and 
isometry group $G=\RR^3 \times SO(3)$ and
the hyperbolic space $H^3$  with  $ R < 0$ and 
isometry group $G=PSL(2,C)$.
The constant curvature spaces $H^3,S^3$ 
will be analysed in detail in the following sections.
The  Euclidean space  $E^3$, being flat, does not lead to a new background, i.e.
the
topological twist is trivial. 

We will
analyse 
the requirements that
an associative 3-cycle $L$ of $SO(3)$ holonomy has to meet 
in order to allow for a solution of the form 
$AdS_2 \times L$, which corresponds to an IR fixed point.
We will see that the cycle has to be of
a constant negative scalar curvature type.
Thus, we will find an $AdS_2 \times ds^2_{H^3}$ solution, 
and its quotients by subgroups of 
the isometry group  $PSL(2,C)$ to make the 3-cycle compact.
Since the spinors are independent of the coordinates on ${H^3}$ these quotient
 solutions are also
supersymmetric. 
We will
construct a regular numerical solution  interpolating between $AdS_5$ in
the UV and  $AdS_2 \times H^3$ in the IR.
The IR fixed point exists at the ``intersection'' of the Coulomb and Higgs branches.
We will analyse the singular supergravity solutions which correspond
to moving into the Higgs and the Coulomb branches.
For negative  constant curvature spaces the singularity will be shown to be
of a ``good'' type
in the  Higgs branch and of a ``bad'' type
in the Coulomb  branch.
This indicates a decoupling of the Higgs branch from the Coulomb branch.
For positive constant curvature spaces 
such as $S^3$ the singularity is of a ``bad'' type
in both the  Higgs and
the Coulomb  branches, which is consistent with the absence of a Higgs branch on the field theory side 
in this case.

The paper is organized as follows.
In section 2 we will discuss
in some detail the twisted theories on the worldvolume of D3-branes wrapping
associative and special 
Lagrangian 3-cycles. We will review the barely $G_2$ manifolds and discuss their
supersymmetric 3-cycles.
In section 3 we will discuss the supergravity equations which have to be solved in
order to provide  supergravity dual descriptions of the twisted D3-branes theories.
In section 4 we will begin by reviewing the possible types
of 3-cycles that should be analysed. We will then
construct solutions to the supergravity equations. We will show
that supersymmetric 
$AdS_2 \times L$ backgrounds are solutions 
provided the associative 3-cycle is of
a constant negative scalar curvature type.
However, we will show that wrapping associative 3-cycles of the type $\Sigma \times S^1$
is  not realized by the $\cN=4$ gauged supergravity. 
We will solve  numerically the supergravity equations and   
construct a regular solution  interpolating between $AdS_5$ in the UV
and  $AdS_2 \times H^3$ in the IR.
We will next analyse the singular supergravity solutions which correspond
to moving into the Higgs and the Coulomb branches and the type of the singularities.
In appendix A we will show how the case of D3-branes wrapping a Riemann surface arises from our 
construction and discuss the nature
of the singularities along the Coulomb and Higgs branches.
In appendix B we will provide some details of the computations for $H^3$ and $S^3$ special
associative 3-cycles.

\section{D3-branes wrapping supersymmetric 3-cycles}

In this section we will discuss two types of supersymmetric 3-cycles, the associative 3-cycle
in $G_2$ holonomy seven-dimensional manifolds and the special Lagrangian 3-cycle
in Calabi-Yau 3-fold. We will review the example of barely $G_2$ manifolds constructed as the 
quotients $(CY_3 \times S^1)/Z_2$, where one can see a relation between the supersymmetric
3-cycles of the seven dimensional manifold and the supersymmetric cycles of the Calabi-Yau
3-fold.
We will analyse the twists associated with D3-branes wrapping the 3-cycles and discuss
the corresponding twisted worldvolume theories.
We note that in the following we will freely use $SU(2)\simeq SO(3)$.

\subsection{Associative 3-cycles}

An associative 3-cycle $L$ is a submanifold of a seven dimensional manifold with $G_2$ holonomy. 
It is a calibrated submanifold with respect to
the associative calibration 3-form $\Phi_{(3)}$ \cite{hl}.
That means that the associative 3-cycles
are volume minimizing in their homology class such that 
\begin{equation}
  \label{eq:asso1}
  \Phi_{(3)} |_L = Vol_L \ .
\end{equation}
They are supersymmetric cycles in the sense that D-branes wrapping them
are supersymmetric configurations \cite{BB}.
Consider Type IIB string theory compactified on a  seven dimensional manifold with $G_2$ holonomy 
$M$, and
a D3-brane wrapping  $L$. 
The brane   configuration is a BPS state of  Type IIB string theory
compactified on $M$. It is a particle in the transverse $R^{2,1}$ directions.
The D3-brane worldvolume theory 
is a twisted field theory. 
The holonomy group of the 3-cycle $L$ is $\subseteq SO(3)$ and its normal bundle $\cN$
 in $M$ 
is $S \otimes V$ \cite{Mclean}, where $S$ is a spin bundle and $V$ a vector bundle.
The four scalars parametrizing the embedding of the D3-brane in
$M$ are sections of the normal bundle of $L$ in $M$ and are therefore not 
ordinary scalars. Let us analyse this structure in more detail.

The  symmetry group of a D3-brane in flat space is
$SO(3,1) \times SO(6)$, where $SO(3,1)$ is the Lorentz symmetry and $SO(6)$ is
the R-symmetry group of the worldvolume theory.
Alternatively, they can be viewed as the symmetry groups of the tangent and normal bundles
respectively.
The worldvolume of the brane is $R \times L$ and the worldvolume theory  is a twisted theory. 
Due to the wrapping the structure group of the normal bundle is broken to $U(1) \times SU(2)_1 \times SU(2)_2$
and 
\begin{equation}
  \label{eq:lbreak}
  SO(3,1) \times SO(6) \rightarrow SO(3)_L \times  U(1) \times SU(2)_1 \times SU(2)_2     \ .
\end{equation}
Here $SO(3)_L$ is the structure group of the tangent bundle of $L$. 

The six scalars parametrizing the embedding of the D3-brane in flat space transform
as  $(\bf{1},\bf{6})$
under $SO(3,1) \times SO(6)$. 
The $\bf{6}$ of the $SO(6)$ 
transforms under $U(1) \times SU(2)_1 \times SU(2)_2$ as
$(\bf{1},\bf{1})_{\bf{\pm 2}} \oplus (\bf{2},\bf{2})_{\bf{0}}$, where the subscript indicates the $U(1)$ charge.
The 
twisting means choosing the holonomy group of the tangent bundle $SO(3)_L$
to be equal to the $SU(2)_1$ part of the structure group of the normal bundle.
After the twisting the six scalars transform as
$(\bf{1},\bf{1})_{\bf{\pm 2}} \oplus (\bf{2},\bf{2})_{\bf{0}}$ under
$U(1) \times SU(2)_L \times SU(2)_2$.
The $(\bf{2},\bf{2})_{\bf{0}}$ parametrize the embedding of
the D3-brane inside the $G_2$ manifold, and
are sections of the normal bundle described above.
The $(\bf{1},\bf{1})_{\bf{\pm 2}}$ are the 
two scalars that parametrize
the embedding of the D3-brane in the 
two
non-compact flat directions normal to the brane.

The sixteen supercharges of the D3-brane in flat space transform
as  $(\bf{2},\bf{1},\bf{4}) \oplus (\bf{1},\bf{2},\bar{\bf{4}})$
under $SO(3,1) \times SO(6)$. 
The $\bf{4}$ of $SO(6)$ transforms under 
$U(1) \times SU(2)_1 \times SU(2)_2$ as 
$(\bf{2},\bf{1})_{\bf{1}}\oplus (\bf{1},\bf{2})_{\bf{-1}}$.
After the twist we remain with 
two covariantly constant supercharges
which transform under 
$U(1) \times SU(2)_L \times SU(2)_2$ as 
$(\bf{1},\bf{1})_{\bf{\pm 1}}$.
The supercharges are ``twisted'' and become scalars rather than spinors.

We will consider $N$ D3-branes wrapping the 3-cycle $L$ in the limit
$l_s \rightarrow 0$ where we keep the volumes of the cycle $L$ and the $G_2$
manifold $M$ fixed \cite{malda}. This is the large volume limit in string units. 
Due to the non-trivial embedding of the cycle in $M$ the worldvolume theory
in this limit is a twisted field theory, as described above. However, note
that in this limit the theory is not sensitive to the global structure of $M$.

While at high energies the worldvolume theory is four-dimensional,
at low energies compared to the inverse size of $L$ the theory is $(0+1)$-dimensional
$U(N)$ gauge theory.
The $(0+1)$-dimensional gauge coupling $g$ is given by
$g^2 = \frac{g_s}{Vol(L)}$.
It has a mass dimension $\frac{3}{2}$.
We will be interested in the infrared (large time)
behaviour of the quantum mechanics.
At low-energies the theory is described by
a supersymmetric $\sigma$-model quantum mechanics with 
{\it two} supercharges. 
The target space has branches which we denote by Higgs and Coulomb,
and is  
not protected from quantum corrections.

The Higgs branch corresponds to the moduli of the associative 3-cycle embedded in the $G_2$
manifold.
The Coulomb branch corresponds to the moduli of the D-branes in the two transverse
non-compact flat directions.
At low energies
we may expect a decoupling of the Higgs and Coulomb branches,
such that the IR physics is that of a supersymmetric $\sigma$-model quantum mechanics 
on the Higgs branch. This is argued for theories with eight supercharges in \cite{A,W}
and  for theories with four supercharges in \cite{KOY}.
We will see that the dual supergravity description indicates
a decoupling of the Higgs and Coulomb branches in our case, which has only two supercharges.

\subsection{Special Lagrangian 3-cycles}

Here we consider Type IIB string theory compactified on a Calabi-Yau 3-fold
$M$. 
Denote by $\kappa$ the Kaehler form of $M$ and by $\Omega$ the holomorphic
$(3,0)$ form.
A 3-cycle $L \in M$ is called special Lagrangian when\\
(i) $\kappa |_L = 0$\\
(ii) $Im(\Omega) |_L = 0$.\\
Consider a D3-brane wrapping  $L$. 
Again, the D3-brane worldvolume theory 
is a twisted field theory. 
The holonomy group of the 3-cycle $L$ is $\subseteq SO(3)$ and its normal bundle $\cN$
 in $M$ can be identified with the cotangent bundle $T^*L$. 
The three scalars parametrizing the embedding of the D3-brane in
$M$ are sections of the normal bundle of $L$ in $M$ and are therefore 1-forms
rather than ordinary scalars. 
Due to the wrapping the symmetry group 
is broken as

\begin{equation}
  \label{eq:lbreak2}
  SO(3,1) \times SO(6) \rightarrow SO(3)_L \times SO(3)_{\cN} \times SO(3)_T  \ .
\end{equation}
Here $SO(3)_L$ is the symmetry group of $L$, 
$SO(3)_{\cN}$ is the structure group of the normal bundle of $L$ in $M$,
$SO(3)_T$ is the symmetry group of the three 
non-compact flat directions transverse to the brane.

The sixteen supercharges of the D3-brane in flat space transform
as  $(\bf{2},\bf{1},\bf{4}) \oplus (\bf{1},\bf{2},\bar{\bf{4}})$
under $SO(3,1) \times SO(6)$. 
Since 
the $\bf{4}$ of $SO(6)$ transforms under a
subgroup $SO(3)_{\cN} \times SO(3)_T$ as $(\bf{2},\bf{2})$,
the supercharges transform under 
$SO(3)_L \times SO(3)_{\cN} \times SO(3)_T$
as $(\bf{2},\bf{2},\bf{2}) \oplus (\bf{2},\bf{2},\bf{2})$.
The 
twisting means choosing the holonomy group of the tangent bundle $SO(3)_L$
to be equal to the structure group of the normal bundle
$SO(3)_{\cN}$.  This  gives four covariantly constant supercharges
which transform as $2(\bf{1},\bf{2})$ under $SO(3)_L \times SO(3)_T$.
The supercharges are ``twisted'' and become scalars rather than spinors.

The six scalars parametrizing the embedding of the D3-brane in flat space transform
as  $(\bf{1},\bf{6})$
under $SO(3,1) \times SO(6)$. 
Since 
the $\bf{6}$ of $SO(6)$ transforms under a
subgroup $SO(3)_{\cN} \times SO(3)_T$ as $(\bf{1},\bf{3}) \oplus (\bf{3},\bf{1})$,
the six scalars  transform under 
$SO(3)_L \times SO(3)_{\cN} \times SO(3)_T$
as $(\bf{1},\bf{1},\bf{3}) \oplus (\bf{1},\bf{3},\bf{1})$.
After the 
twist
they transform as $(\bf{1},\bf{3}) \oplus (\bf{3},\bf{1})$
under $SO(3)_L \times SO(3)_T$.
The $(\bf{1},\bf{3})$ are the three scalars that parametrize
the embedding of the D3-brane in the 
 three 
non-compact flat directions normal to the brane.
The $(\bf{3},\bf{1})$ correspond to the 1-form that parametrizes
the embedding of the D3-brane in the 
directions normal to the brane in $M$.

We would like to consider $N$ D3-branes wrapping the 3-cycle $L$ in the above limit.
At low-energies the theory is described by
a supersymmetric $\sigma$-model quantum mechanics with 
{\it four} supercharges. 
The Higgs branch corresponds to the moduli of the special Lagrangian cycle embedded in the Calabi-Yau
3-fold. The Coulomb branch corresponds to the moduli of the D-branes in the three transverse
non-compact flat directions.
Again, we may expect 
a decoupling of the Higgs and Coulomb branches in the IR.

\subsection{Barely $G_2$ manifolds}

Let us illustrate the associative and special Lagrangian 3-cycles by discussing the
barely $G_2$ manifolds \cite{harvey}. These are constructed as 
\begin{equation}
  \label{eq:g2cy3}
  G=\left(CY_3 \times S^1 \right)/ Z_2,
\end{equation}
where $CY_3$ is the Calabi-Yau 3-fold and $Z_2$ acts as 
$(\sigma,-1)$. $\sigma$ is an anti-holomorphic 
involution on the Calabi-Yau 3-fold. Its action
on the Kaehler form $\kappa$ of $CY_3$, and on the holomorphic
$(3,0)$ form  $\Omega$ is
\begin{equation}
\sigma^{*}(\kappa) = -\kappa, ~~~~~\sigma^{*}(\Omega) = \bar{\Omega} \ .
\end{equation}
There is only one covariantly constant spinor on $G$, which corresponds to the linear combination
of the two covariantly constant spinors with opposite chirality on $CY_3$.

The associative calibration 3-form reads \cite{harvey} 
\begin{equation}
  \label{eq:caliform}
  \Phi_{(3)}=J \wedge e^1 +Re[\Omega],
\end{equation}
where $e^1$ is a one-form cotangent to the circle $S^1$.

There are two types of supersymmetric associative 3-cycles ${\cal C}_3$.
The first one is constructed as a 
product of $C\times S^1$ where $C$ is
a holomorphic cycle in $CY_3$, modded by the action of $Z_2$
\begin{equation}
  \label{eq:cycle1}
  {\cal C}_3 = \left(C \times S^1 \right)/ Z_2 \ .
\end{equation}
Note that $\sigma$ maps $C$ to $-C$. 

The second  type is constructed by the quotient of 
a special Lagrangian cycle $L$ in $CY_3$
\begin{equation}
  \label{eq:cycle2}
  {\cal C}_3=L/Z_2.
\end{equation}
For this type of associative 3-cycles the calibration form is just $Re[\Omega]$ as for
a special Lagrangian 3-cycle.
Note, however, that 
the ambient space in which the 3-cycle is
embedded is different, and the number of supersymmetries preserved by the associative 3-cycle
is half that of the special Lagrangian one.

\section{The supergravity equations}

We aim at a dual supergravity description
of the 
worldvolume theory of $N$ D3-branes wrapped around a supersymmetric 3-cycle 
$L$.
As discussed above,
the worldvolume  theory is
twisted, and the twist induces the boundary conditions 
on the gauge field coming from the metric upon compactification.
It also implies that a scalar  operator in the ${\bf 20}$ representation of $SO(6)$
is  turned on, as in \cite{malda}. 

We will use a five-dimensional truncation of the full type IIB supergravity. 
To find a suitable truncation we observe that the holonomy of the $L$ is 
$SO(3)$ so for the twist we will need an $SO(3)$ gauge potential. 
One suitable supergravity theory in five dimensions 
is the $\cN=4$ $SU(2) \times U(1)$ gauged supergravity described in \cite{romans}
 which was shown to be a consistent truncation of type IIB supergravity in \cite{pope}. 
The embedding of this five-dimensional supergravity in type IIB supergravity
is done via the breaking $SO(6) \rightarrow U(1) \times SU(2)_1 \times SU(2)_2$,
where $SU(2) \times U(1)$ of the $\cN=4$ is the  $SU(2)_1 \times U(1)$.
From the discussion of the previous section we know that this can realize the twist required
for D3-branes wrapping an associative 3-cycle.

We are looking for a supersymmetric solution of the theory of \cite{romans}.
To this end we have to check the supersymmetric
 variations of the fermionic fields. These are given
by
\begin{equation}
  \label{eq:vargravi}
  \delta \psi_{\mu a}\;=\; D_{\mu}\, \varepsilon_a\, +\,i\, 
\gamma_{\mu}\,T \left(\Gamma_{45}\right)_{ab}\varepsilon^b\;+
\,i\;\frac{1}{6}\;\sqrt{\, \frac{1}{2}}\, 
\left(\,\gamma_{\mu}^{\;\nu \rho}\;-\;4\,
\delta_{\mu}^{\nu}\,\gamma^{\rho}\,\right) 
\left(\,H_{\,\nu \rho a b}\;+\; \sqrt{\,\frac{1}{2}}\,h_{\,\nu \rho a b}\, \right)\varepsilon^b,
\end{equation}
for the four gravitini and
\begin{equation}
  \label{eq:varsp2}
  \delta \chi_a\;=\;-i\,\sqrt{\,\frac{1}{2}}\,\gamma^{\mu}\,(\partial_{\mu}\, 
\Phi)\,\varepsilon_a\;+\;A(\Gamma_{45})_{ab}\, \varepsilon^b\;+\;\frac{1}{2}\,
\sqrt{\,\frac{1}{6}}\,\gamma^{\, \mu\nu}\,\left(\,H_{\,\mu \nu a b}\;-\;\sqrt{\,2}\,h_{\,\mu \nu a b}\,\right)\,\varepsilon^b
\end{equation}
for the remaining four spin-$\frac{1}{2}$ fields. The notation agrees with \cite{romans} except that we use a space-time metric with signature $(-,+,+,+,+)$. The definitions are
\begin{eqnarray}
  \label{eq:defis}
  A &  \;\equiv & \;\frac{1}{2}\, \sqrt{\,\frac{1}{6}}\,g_2\, \xi^{-1}\;-\;\frac{1}{2}\, 
\sqrt{\,\frac{1}{3}}\,g_1\,\xi^{2}, \\
  T &  \;\equiv & \;\frac{1}{6}\, 
\sqrt{\,\frac{1}{2}}\,g_2\, \xi^{-1}\;+\;\frac{1}{12}\,g_1\,\xi^{2}, \\
  h_{\,\mu \nu a b} & \;\equiv & \;-\xi^{-2}\, \Omega_{ab}\, f_{\mu \nu}, \\
  H_{\,\mu \nu a b} & \;\equiv & \; \xi \,(\, F_{\mu \nu}^I \,
\left( \Gamma_I \right)_{ab} \;+ 
\;B_{\mu \nu}^{\alpha} \left(\Gamma_{\alpha} \right)_{ab}), \\
  \xi & \;\equiv & \;\exp{\,\sqrt{\,\frac{2}{3}}}\, \Phi \ .
\end{eqnarray}
The $\gamma_{\nu}, \nu=0,...,4$ are the five dimensional space-time gamma matrices, whereas the 
$\Gamma_{I,\alpha}$ are the $Spin(5)$ gamma matrices with the index ranges  $I=1,2,3$ 
and $\alpha = 4,5$. The $a,b$ are $Spin(5)$ spinor indices and $\Omega_{ab}$ is the 
antisymmetric matrix used to
lower
 these indices. $F_{\mu \nu}^I$ and $f_{\mu \nu}$ are the $SU(2)$ and $U(1)$ gauge 
fields respectively, $g_2$ and $g_1$ are the corresponding coupling constants, $\Phi$ 
is a scalar field 
and $B_{\mu \nu}^{\alpha}$ are antisymmetric tensor fields transforming as a $U(1)$ doublet. 

The $SU(2) \times U(1)$ is embedded in $Spin(5)$ by 
taking $\Gamma_{I45}$ and $\Gamma_{45}$ as the respective generators. 
We therefore have the covariant derivative on the spinors $\varepsilon_a$
\begin{equation}
  \label{eq:covder1}
  D_{\mu} \,  \varepsilon_a\;=\; \nabla_{\mu} \,\varepsilon_a \;+\;\frac{1}{2}\,g_1 \,a_{\mu} 
(\Gamma_{45})^{\,b}_a \,\varepsilon_b \;+ \; \frac{1}{2}\,g_2 
\,A_{\mu}^I (\Gamma_{I45})^{\,b}_a \,\varepsilon_b,
\end{equation}
with $\nabla_{\mu}$ denoting the usual space-time covariant 
derivative and $a_{\mu},A_{\mu}^I$ are
the $U(1), SU(2)$ gauge potentials.
Furthermore we will use the freedom to rescale the coupling constants as in \cite{romans} and define $\bar{g} \equiv \sqrt{2}g_1, g_2 \equiv \bar{g}$. 
Note that we cannot set the $U(1)$ coupling $g_1$ simply to zero since the kinetic term of the 
 $B_{\mu \nu}^{\alpha}$ field goes like $\frac{1}{g_1}$  \cite{romans}
\footnote{A supergravity action where the $U(1)$ coupling can
be taken to zero has been constructed in \cite{Cowdall:1999rs}.}.

For later reference it is useful to recall
the five-dimensional
field equations for zero fermionic fields
which follow from the Lagrangian given in \cite{romans}. In terms of our conventions and with $a_{\mu},B_{\mu \nu}^{\alpha}$ set to zero we obtain for the metric
\begin{equation}
  \label{eq:eomg}
  \cR_{\mu \nu}-3\partial_{\mu}\varphi\partial_{\nu}\varphi-2e^{2\varphi}F^{I}_{\mu \alpha}F_{\nu\;I}^{\alpha}+\frac{1}{3}e^{2\varphi}F^2g_{\mu \nu}+\frac{4}{3}P\left(\varphi\right)g_{\mu \nu}\,=\,0, 
\end{equation}
and for the scalar
\begin{equation}
  \label{eq:eoms}
  \nabla^2 \varphi-\frac{1}{3}e^{2\varphi}F^2 +\frac{2}{3} 
\frac{\partial P\left(\varphi\right)}{\partial \varphi}\,=\,0 \ ,
\end{equation}
where for notational simplicity we 
 introduced $\varphi \equiv \sqrt{\frac{2}{3}}\Phi$ 
and used $F^2$ as short for $F^{I}_{\mu \nu}F_{I}^{\mu \nu}$. 
The scalar potential $P\left(\varphi\right)$ reads
\begin{equation}
  \label{eq:pscalar}
  P\left(\varphi\right)\,=\,\frac{1}{8}\bar{g}^2\left(e^{-2\varphi}+2e^{\varphi}\right) \ .
\end{equation}

For a supersymmetric solution the 
variations (\ref{eq:vargravi}),(\ref{eq:varsp2}) of the fermionic fields must be zero. 
As the five dimensional metric we take the ansatz
\begin{equation}
  \label{eq:metric}
    ds^2=e^{2f}(-dt^2+dr^2) +e^{2g} ds^2_L \ ,
\label{metric}
\end{equation}
where by $ds^2_L$ we denote the metric on the 3-cycle $L$
\begin{equation}
  \label{eq:genmetric}
  ds_L^2= g_{ij}dx^idx^j \ .
\label{SL}
\end{equation}

We assume the functions $f,g$ to depend only on the radial coordinate $r$.
We also take the scalar field 
$\varphi$ 
to depend only on $r$.
$\varphi$ is dual to the turned on scalar operator.

Thus, we have to solve the following equations
\begin{eqnarray}
  \label{eq:newvars}
   & & \delta \psi_{\mu a}\,=\, \partial_{\mu}\, 
\varepsilon_a\,+\,\frac{1}{4}\omega_{\mu}^{ij}\gamma_{ij}\, 
\varepsilon_a\,+\,\frac{1}{2}\bar{g}A_{\mu}^{I}\left(\Gamma_{I45}\right)_a^{\;b}\, 
\varepsilon_b\, - \\ \nonumber
 & &-\,i\, \gamma_{\mu}\,T \left(\Gamma_{45}\right)_a^{\;b}\varepsilon_b\;-\,i\;
\frac{1}{6}\;\sqrt{\, \frac{1}{2}}\, \left(\,\gamma_{\mu}^{\;\nu \rho}\;-\;4\,
\delta_{\mu}^{\nu}\,\gamma^{\rho}\,\right) e^{\varphi} F_{\nu \rho}^I
\left(\Gamma_I\right)_a^{\;b}\varepsilon_b\,=\,0,
\end{eqnarray}
\begin{eqnarray}
  \label{eq:newvars2}
  \delta \chi_a\;=\;-i\,\frac{\sqrt{3}}{2}\,\gamma^r\,(\partial_r\, \varphi)\,
\varepsilon_a\;-\;A(\Gamma_{45})_a^{\;b}\, \varepsilon_b\;-\;\frac{1}{2}\,
\sqrt{\,\frac{1}{6}}\,\gamma^{\, \mu\nu}e^{\varphi}F_{\mu \nu}^I \left(\Gamma_I\right)_a^{\;b}\,\varepsilon_b\,=\,0,
\end{eqnarray}
with the definitions
\begin{eqnarray}
  \label{eq:supersach}
  & & A\,\equiv\,\frac{1}{2}\sqrt{\,\frac{1}{6}}\bar{g}
\left(e^{-\varphi}-e^{2\varphi}\right), \\ \nonumber
  & & T\,\equiv\, \frac{1}{12\sqrt{2}}\bar{g}\left(2e^{-\varphi}+e^{2\varphi}\right).
\end{eqnarray}

We note that for a time independent spinor $\varepsilon_a$ we can combine the $t$-component of equations (\ref{eq:newvars}) and (\ref{eq:newvars2}) to get the relation  
\begin{equation}
  \label{eq:simplerel}
  f^{\prime}\,-\,\varphi^{\prime}\,=\,\pm\frac{\bar{g}}{2\sqrt{2}}e^f e^{2\varphi},
\end{equation}
for $\left(\Gamma_{45}\right)_a^{\;b}\varepsilon_b\,=\,\mp i\varepsilon_a$. This relation is independent of the metric of $L$.

At the boundary of $AdS_5$ (small $r$) we impose the metric
\be
ds^2 \sim \frac{-dt^2+ dr^2+  ds^2_L }{r^2} \ .
\ee
This implies at small $r$ 
the boundary conditions $f(r), g(r) \sim -Log(r)$.
Note that we work in the units where the $AdS_5$ radius is one.
In order to restore the units and the $N$ dependence we have to multiply
the metric by $R^2_{AdS} = \sqrt{4 \pi g_s N}\alpha'$.

We impose the condition   
\begin{eqnarray}
\label{eq:susubr}
  \gamma_1\, \varepsilon_a & = & \varepsilon_a,
\end{eqnarray}
where the subscripts are flat indices associated to the 
space-time indices according to (\ref{eq:mv1}). 
This condition
is met by (super)covariantly 
constant spinors on $AdS$ which depend only on 
$r$ \cite{Lu:1997rh,pope2}.

To realize the twist 
we have to satisfy
\begin{equation}
  \label{eq:twist1}
  \frac{1}{4}\omega_{\hat{\mu}}^{\hat{i}\hat{j}}\gamma_{\hat{i}\hat{j}}\, 
\varepsilon_a\,=\,-\frac{1}{2}\bar{g}A_{\hat{\mu}}^{I}\left(\Gamma_{I45}\right)_a^{\;b}\, 
\varepsilon_b,
\end{equation}
where the hatted indices are 
(${\hat{\mu}}$ curved, ${\hat i}$ flat) 
indices along $L$. We do this by appropriately mapping the generators of the holonomy group to the generators of the $SU(2)$ gauge group and letting the $U(1)$ generator $\Gamma_{45}$ act on the spinors as $\left(\Gamma_{45}\right)_a^b\,\varepsilon_b\,=\,\pm i \varepsilon_a$. 
The boundary conditions on the metric imply that the positive sign is 
the correct choice as can be seen from the gravitino variation (\ref{eq:vargravi}) in $t$ direction. 
By equating the spin connection with the gauge connection we thus get
\begin{equation}
  \label{eq:twist2}
  \gamma_{\hat{i}\hat{j}}\varepsilon_a \propto i\left(\Gamma_{f(\hat{i},\hat{j})}\right)^{\;b}_a\varepsilon_b,
\end{equation}
where $f$ establishes the map between the generators of the two groups (see below and appendix).


\section{Supergravity backgrounds}

In this section we will  construct analytical and numerical
solutions to the supergravity equations.

\subsection{Three manifolds}

In the following we discuss what types of 3-cycles we should consider.
Thurston introduced eight geometric types in the classification of closed 3-manifolds \cite{t},
which we will briefly review
\footnote{For the relation between Thurston classification and the classification of spatially
homogeneous metrics in relativistic cosmology see \cite{F}.}. 
The main points of the classification
are that there are only eight basic homogeneous geometries, up to an equivalence 
relation, that can be supported by closed 3-manifolds and that if a closed 3-manifold
of a given topology admits one of these geometric types then it is unique.

One considers an orientable, connected, complete and  
simply connected Riemannian 3-manifold $X$ which is homogeneous with respect to an
orientation preserving group of isometries $G$.
The eight geometric types classify $(X,G)$. 
The equivalence relation $(X,G) \sim (X',G')$ holds when
there is a diffeomorphism of $X$ onto $X'$ which takes the action of $G$ onto
the action of $G'$.
Out of these types one constructs spaces 
(geometric structures)
$M \simeq X/\Gamma$ where $\Gamma$ is a subgroup of $G$. Here the action of $\Gamma$
is discontinuous, discrete and free. 
$M$ is locally homogeneous with respect to the metric on $(X,G)$ \footnote{
$M$ is called locally homogeneous if for any two points in $M$ there are neighborhoods
of these two points and an isometry that maps them to each other.}, it is
isometric to the quotient of $X$ by $\Gamma$.
 
The first three types in the classification
are based on the three constant curvature spaces, the 3-sphere $S^3$
which has a positive scalar curvature $ R >0$
and isometry group $G=SO(4)$, the Euclidean space  $E^3$ with  $ R = 0$ and 
isometry group $G=\RR^3 \times SO(3)$ and
the hyperbolic space $H^3$  with  $R < 0$ and 
isometry group $G=PSL(2,C)$.
The constant curvature spaces $S^3,H^3$ will be analysed in detail in the following sections.
The  Euclidean space  $E^3$, being flat, does not lead to a new background, i.e.
the
topological twist is trivial. This class includes, for instance, the 3-torus $T^3$.

The next two types are based on $S^2 \times E^1$ and  $H^2 \times E^1$.
In this class we have, for instance, $\Sigma_g \times S^1$ where $\Sigma_g$ is a genus $g$ Riemann surface.

Type six is the universal cover of the fiber bundle whose base is $H^2$ 
and whose fiber is spanned by tangent vectors of unit length. The last two types are the
Heisenberg group and the solvable Lie group.
These types are anisotropic geometries and one can write an explicit metric for them \cite{F}.
However, it is not clear whether any of these last three types
can be realized as a supersymmetric  3-cycle.

\subsection{$AdS_2\times L$ solutions}

In this subsection we analyse
the requirements an associative 3-cycle with $SO(3)$ holonomy 
has to meet in order to allow for a solution of the form 
$AdS_2 \times L$, which corresponds to an IR fixed point.
We will see that for our ansatz $L$ has to be of constant negative
curvature type.
We consider the ansatz
(\ref{metric}),(\ref{SL}).
Due to the twist we can establish a
relation between the $SU(2)$ gauge field strength and the curvature of $L$
\begin{equation}
  \label{eq:gfcurv}
 F_{\mu \nu}^{\;\;I} \left(\Gamma_I\right)_a^{\;b}\,\varepsilon_b\,=\,i\frac{1}{2\bar{g}}R_{\mu \nu}^{\;\;\;ij}\,\gamma_{ij}\,\varepsilon_a,
\end{equation}
which follows from (\ref{eq:twist1}). We now take
 the variation of the gravitino (\ref{eq:newvars}) which is polarized in the $t$ direction. This may be written as 
\begin{equation}
  \label{eq:tvary}
  \left(\frac{1}{2}f^{\prime}\,+\, e^f\,\frac{1}{12\sqrt{2}}\bar{g}
\left(2e^{-\varphi}+e^{2\varphi}\right)\,+\,\frac{1}{6} \sqrt{\frac{1}{2}}e^fe^{\varphi}\frac{1}{2\bar{g}}\gamma^{\nu \rho}R_{\nu \rho}^{ij}\gamma_{ij}\right)\gamma_0\,\varepsilon_a=0,
\end{equation}
so that
\begin{equation}
  \label{eq:lhsrhs}
  \left(\frac{1}{2}f^{\prime}\,+\, e^f\,\frac{1}{12\sqrt{2}}\bar{g}
\left(2e^{-\varphi}+e^{2\varphi}\right)\right)\gamma_0\,\varepsilon_a \,=\,-\frac{1}{6} \sqrt{\frac{1}{2}}e^fe^{\varphi}\frac{1}{2\bar{g}}\gamma^{\nu \rho}R_{\nu \rho}^{\;\;\;ij}\gamma_{ij}\gamma_0\,\varepsilon_a.
\end{equation}
The lhs of (\ref{eq:lhsrhs}) depends only on $r$ so the rhs must be independent of the coordinates on $L$. The rhs depends on the coordinates on $L$ through the term $\gamma^{\nu \rho}R_{\nu \rho}^{\;\;\;ij}\gamma_{ij}$ so that we have the condition
\begin{equation}
  \label{eq:Rconst}
  \gamma^{\nu \rho}R_{\nu \rho}^{\;\;\;ij}\gamma_{ij}\,=\,const \ ,
\end{equation}
up to an $r$ dependent term coming from the vielbein needed to change the
curved indices in $ \gamma^{\nu \rho}$ to flat ones. 
In three dimensions the Weyl tensor vanishes identically so that the Riemann tensor can be expressed in terms of Ricci tensor and scalar curvature 
\begin{equation}
  \label{eq:Riemann}
  R_{\kappa \lambda \mu \nu}\,=\,R_{\kappa \mu}g_{\lambda \nu}-R_{\lambda \mu}g_{\kappa \nu}+R_{\lambda \nu}g_{\kappa \mu}-R_{\kappa \nu}g_{\lambda \mu}-\frac{R}{2}\left(g_{\kappa \mu}g_{\lambda \nu}-g_{\kappa \nu}g_{\lambda \mu}\right).
\end{equation}
Using this in (\ref{eq:Rconst}) we see that the scalar curvature has to be constant.  For cycles with $SO(3)$ holonomy this means that they are maximally symmetric cycles which satisfy
\begin{equation}
  \label{eq:maxsymm}
  R_{\kappa \lambda \mu \nu}\,=\,\frac{R}{6}\left(g_{\kappa \mu}g_{\lambda \nu}-g_{\kappa \nu}g_{\lambda \mu}\right).
\end{equation}

We are now in the position to investigate (supersymmetric) solutions of the equations (\ref{eq:newvars}) and (\ref{eq:newvars2}) which are of the form $AdS_2 \times L$. 
We require 
\begin{equation}
  \label{eq:bc11}
  e^{2f}\,=\,\frac{A^2}{r^2},
\end{equation}
according to the ansatz (\ref{eq:metric}). We rewrite (\ref{eq:tvary}) for the maximally symmetric cycle by using (\ref{eq:bc11}). This leads to 
\begin{equation}
  \label{eq:fterm}
  \frac{1}{6\bar{g}}A\sqrt{\frac{1}{2}}e^{-2g}R\,=\,-\frac{1}{2}e^{-\varphi}+A\frac{\bar{g}}{12\sqrt{2}}\left(2e^{-2\varphi}+e^{\varphi}\right).
\end{equation}
In order to satisfy equation (\ref{eq:simplerel}) with the boundary condition (\ref{eq:bc11}) $\varphi$ has to approach a constant value in the far IR ($r\rightarrow \infty$). Equation (\ref{eq:fterm}) then also requires $e^{2g}$ to be asymptotically constant. So supersymmetric solutions in the IR with an $AdS_2$ part are indeed of the product form $AdS_2 \times L$ where $L$ has constant size.

We can plug this additional information into the remaining fermion variations of (\ref{eq:newvars}) and (\ref{eq:newvars2}). This results in
\begin{equation}
  \label{eq:R1}
  R\,=\,-\frac{1}{2}e^{2g}\bar{g}^2\left(2e^{-2\varphi}+e^{\varphi}\right),
\end{equation}
and
\begin{equation}
  \label{eq:R2}
  R\,=\,e^{2g}\bar{g}^2\left(e^{-2\varphi}-e^{\varphi}\right).
\end{equation}
Combining these two equations and using (\ref{eq:fterm}) we find 
\begin{equation}
  \label{eq:unique}
  e^{3\varphi}=4,\:\:e^{2g}= -\frac{1}{3\bar{g}^2}4^{\frac{2}{3}}R
,\:\:e^{2f}=\frac{4^{\frac{2}{3}}}{2\bar{g}^2r^2} \ ,
\end{equation}
where  $R$ is the scalar curvature of $L$. From (\ref{eq:R1}) and (\ref{eq:unique}) we
see that the the scalar curvature of $L$ has to be negative.
We conclude that for a supersymmetric solution
of the type $AdS_2 \times L$  the 3-cycle $L$ has to be
of negative constant curvature type.
The solution preserves two supercharges.
 
Thus, there is an $AdS_2 \times ds^2_{H^3}$ solution and we can take its quotients by
the isometry group  $PSL(2,C)$ to make the 3-cycle compact.
Since the spinors are independent of the coordinates on ${H^3}$ these quotient
 solutions are 
supersymmetric too.
If we take the $ds_{H^3}^2$ line element 

\begin{equation}
ds_{H^3}^2 = 
d\phi ^2 + \sinh^2 \phi d \theta ^2 + \sinh^2 \phi \sin^2\theta d \nu^2 \ ,
\end{equation}
we get 
\begin{equation}
  \label{eq:csol1}
  e^{3\varphi}=4,\:\:e^{2g}=4^{-\frac{1}{3}},\:\:e^{2f}=\frac{1}{4^{\frac{4}{3}}r^2},
\end{equation}
where we have set $\bar{g}=2\sqrt{2}$.

Using  \cite{pope} we can lift the five-dimensional background
to ten dimensions.
The ten-dimensional metric reads
\begin{eqnarray}
  \label{eq:metric10}
  & & ds_{10}^2=\sqrt{\Delta}\left(\frac{1}{4^{\frac{4}{3}}r^2}
ds_2^2+4^{-\frac{1}{3}}ds^2_{H^3}\right)+ 
\\ \nonumber
  & & +\frac{1}{8}2^{\frac{1}{3}}\sqrt{\Delta}d\xi^2+
\frac{1}{8}2^{-\frac{1}{3}}\Delta^{-\frac{1}{2}}\sin^2\xi d\tau^2+ \\ \nonumber
  & & +\frac{1}{8}2^{-\frac{1}{3}}
\Delta^{-\frac{1}{2}}\cos^2\xi\left\{\left(\sigma_1-\cosh\phi d\theta \right)^2 + 
\left(\sigma_2-\cos\theta d\nu \right)^2 + 
\left(\sigma_3+\sin \theta \cosh\phi d\nu \right)^2 \right\},
\end{eqnarray}
where $(\xi,\tau,\alpha,\beta,\gamma)$ parametrize 
the compactifying $S^5$ and the 1-forms $\sigma_i$ are 
related to the angles $(\alpha,\beta,\gamma)$ by $\sigma_1+
i\sigma_2=e^{-i\gamma}(d\alpha+i \sin\alpha d\beta)$ and $\sigma_3=d\gamma+\cos \alpha d\beta$. Furthermore
\begin{equation}
  \label{eq:ddelta}
  \Delta=4^{\frac{2}{3}} \sin^2 \xi + 4^{-\frac{1}{3}} \cos^2 \xi,
\end{equation}
where $ds_2^2=-dt^2+dr^2$.

It is illuminating 
take a closer a look at the field equations 
(\ref{eq:eomg}) and  (\ref{eq:eoms}). 
From (\ref{eq:eomg}) we get for the Ricci Scalar $\cR$ of the five-dimensional
background
\begin{equation}
  \label{eq:scurv}
  \cR\,=\,3\left(\partial \varphi\right)^2+\frac{1}{3}e^{2\varphi}F^2-\frac{20}{3}P\left(\varphi \right).
\end{equation}
Consider solutions where  $\varphi$ is constant.
Equation
 (\ref{eq:eoms}) reads now
\begin{equation}
  \label{eq:eoms1}
e^{2\varphi}F^2 - 2 
\frac{\partial P\left(\varphi\right)}{\partial \varphi}\,=\,0 \ ,
\end{equation}
and
implies  that $F^2$ is constant.
Equation (\ref{eq:scurv}) and (\ref{eq:pscalar})
imply then that the five-dimensional scalar curvature
$\cR$ is constant and negative.
Since $F^2$ can be written as a product of $e^{-4g}$ times some function of the coordinates on $L$ 
we deduce that both factors have to be constant. 
Consider $AdS_2 \times L$ solutions.
Then the above implies that $L$ is a constant curvature 3-fold.

A solution of the form $AdS_2 \times S^3$  is excluded at the level of the field equations (\ref{eq:eomg}) and  (\ref{eq:eoms}). 
Combining the two field equations and taking $\varphi=const.$ we get for the five dimensional Ricci tensor
\begin{equation}
  \label{eq:ricci5d}
  \cR_{\mu \nu}\,=\,\frac{4}{\bar{g}^2}e^{2\varphi}\hat{R}^{-4}g_{\mu \nu}-\frac{1}{2}\bar{g}^2e^{\varphi}g_{\mu \nu},
\end{equation}
where $\hat{R}=e^{g}$ denotes the radius of the $S^3$. 
Note that the term in (\ref{eq:ricci5d}) proportional to $\hat{R}^{-4}$ 
vanishes for the components of
$\cR_{\mu \nu}$ which are not along
the sphere.
The time component of (\ref{eq:ricci5d}) fixes the radius of the $AdS_2$-part in terms of $\varphi$ for the ansatz (\ref{eq:bc11}) as
\begin{equation}
  \label{eq:Rads}
  A^2\,=\,\frac{2}{\bar{g}^2}e^{-\varphi}.
\end{equation}
On the other hand we can use (\ref{eq:scurv}) and (\ref{eq:eoms1}) 
to express $\cR$ 
in terms of $\varphi$ 
\begin{equation}
  \label{eq:TCurv}
  \cR\,=\,-\frac{3}{2}\bar{g}^2e^{\varphi}-\bar{g}^2e^{-2\varphi}.
\end{equation}
For an $AdS_2 \times S^3$ background we have
$\cR\,=\,-\frac{2}{A^2}+\frac{6}{\hat{R}^2}$,
which upon using (\ref{eq:Rads}) leads 
to a contradiction.

Consider now 
3-cycles based on the types
$S^2 \times E^1$ and  $H^2 \times E^1$. They too have
 a constant curvature and one may expect
that they  will
give rise to $AdS_2$ solutions of the gauged supergravity. 
However, in this case the ansatz (\ref{eq:metric}) does not allow for a supersymmetric solution.  
The reason is that from the variation of the gravitino (\ref{eq:newvars}) polarized along the direction of $E^1$ we get for the ansatz (\ref{eq:metric}) one more relation between $g$ and $f$ in addition to 
equations (\ref{eq:sincy}). This prevents a solution.  
The nontrivial part of the spin connection of these cycles comes from the two dimensional part, i.e.
$S^2$ or $H^2$. Therefore they have reduced holonomy.
To implement the twist
we need an abelian truncation of the $SU(2) \times U(1)$  gauged supergravity. 
That means that for such configurations only the $U(1) \subset SU(2)$ is excited.
This  breaks less supersymmetry compared to the cycles with $SU(2)$ holonomy. For an appropriate ansatz there are solutions of this truncated gauged supergravity which then do not correspond to D3-branes wrapped on associative 3-cycles in $G_2$ 
holonomy manifolds but rather to configurations of D3-branes wrapping around a Riemann surface
$\Sigma$  in $CY$ \cite{malda}. 
The details of this calculation are given in appendix A. 
From \cite{pope} it is clear that  from the ten-dimensional point of 
view the abelian truncation used here coincides with the truncation used in \cite{malda} to study D3-branes
wrapped around a Riemann surface in a Calabi-Yau 3-fold.

We conclude that the $\cN=4$ five-dimensional supergravity that we are using is not capable
of describing the wrapping of associative 3-cycle of the type $\Sigma \times S^1$.
Consulting section 2.3, we see that what is missing in that context is the implementation
of the $Z_2$ quotient which truncates the supersymmetry by a factor of two.
A similar situation arises in the analysis of  \cite{g2} for M5-branes wrapped around associative 3-cycles.

In section 2, we described a case (\ref{eq:g2cy3}),
where the associative 3-cycle corresponds to a special Lagrangian 3-cycle.
It is natural to ask whether we can argue based on the $AdS_2 \times L$ solution 
for an associative 3-cycle $L$, that there should be an $AdS_2 \times L$ solution for the
corresponding special Lagrangian
3-cycle. Of course, the latter solution should preserve twice as much supercharges.
For comparison we can look at the example of a Calabi-Yau 3-fold constructed
as $(K3 \times T^2)/\Theta$ \cite{fer}, where the $\Theta$ is a freely acting involution
on $K3$ and an inversion on $T^2$.
In this case we have supersymmetric 2-cycles (Riemann surfaces) in the Calabi-Yau 3-fold that correspond
to Riemann surfaces in $K3$.
D3-branes wrapped on a Riemann surface in the Calabi-Yau 3-fold have an $AdS_3$ solution that preserves 
four supercharges. However, there is no such $AdS_3$ solutions that preserves eight supercharges
that corresponds to  D3-branes wrapped on a Riemann surface in $K3$ \cite{malda}.
Therefore it seems that we cannot draw a conclusion from the 
$AdS_2$ solutions in the associative 3-cycle case, as to the existence
of $AdS_2$ solutions in the special Lagrangian case.

\subsection{$D3$-brane wrapped on $H^3$}

Consider the $N$ $D3$-branes wrapped on an associative 3-cycle 
$H^3$.
Recall that for the five dimensional metric we take the ansatz
\begin{equation}
  \label{eq:hmetric}
   ds^2=e^{2f}(-dt^2+dr^2) +e^{2g}(d\phi ^2 + \sinh^2 \phi d \theta ^2 + \sinh^2 \phi \sin^2\theta d \nu^2),
\end{equation}
where $f,g$ depend only on $r$. 
The second part of the above metric is the metric of the $H^3$.
We have
 to solve (\ref{eq:newvars}) and (\ref{eq:newvars2}) in this background.

The twist implies by (\ref{eq:gp1}) that 
the $SU(2)$ gauge potentials are 
\begin{eqnarray}
  \label{eq:gpnew}
  & & A_{\theta}^{1}=a\cosh \phi, \\ \nonumber
  & & A_{\nu}^{2}=b\cos \theta, \\ \nonumber
  & & A_{\nu}^{3}= c\sin \theta \cosh \phi.
\end{eqnarray}

In order to take advantage of the conditions (\ref{eq:susubr}) 
we write (\ref{eq:newvars}) more explicitly, as done in appendix B in equations
(\ref{eq:expl1}), (\ref{eq:explphi}), (\ref{eq:expltheta}) and (\ref{eq:explnu}).

For the solution we impose on the field strengths (\ref{eq:gf1}) 
\begin{equation}
  \label{eq:c1}
  F^{2}_{\nu \theta} \propto \sin \theta \sinh^2 \phi\; \rightarrow \; \frac{ac}{b}\,=\,-\frac{1}{\bar{g}} 
\end{equation}
and
\begin{equation}
  \label{eq:c2}
  F^{3}_{\nu \theta}=0\; \rightarrow \; c=-ab\bar{g}.
\end{equation}
We solve these constraints by setting
\begin{equation}
  \label{eq:cs1}
  a\,=\, \frac{1}{\bar{g}}\;,\;  b\,=\, \frac{1}{\bar{g}}\;,\;  c\,=\, -\frac{1}{\bar{g}}.
\end{equation}

From the twist we get the following constraints on the Killing spinors 
\begin{eqnarray}
  \label{eq:kc1}
  \gamma_{23} \varepsilon_a\,=\, \pm i \left(\Gamma_1 \right)_a^{\;b} \varepsilon_b \\ \nonumber
  \gamma_{24} \varepsilon_a\,=\, \mp i \left(\Gamma_3 \right)_a^{\;b} \varepsilon_b \\ \nonumber
  \gamma_{34} \varepsilon_a\,=\, \pm i \left(\Gamma_2 \right)_a^{\;b} \varepsilon_b,
\end{eqnarray}
for $\left(\Gamma_{45}\right)_a^{\;b}\varepsilon_b\,=\,\pm i\varepsilon_a$ 
and where we have used (\ref{eq:cs1}).

Given the above constraints the supersymmetric variations vanish for
\begin{eqnarray}
  \label{eq:sol3}
  & & g^{\prime}\,=\,-e^f \left\{-\sqrt{2}e^{\varphi}e^{-2g}\frac{1}{\bar{g}}\,+\, \frac{\bar{g}}{6\sqrt{2}}\left(2e^{-\varphi}\,+\,e^{2\varphi}\right)\right\}, \\ \nonumber
  & & f^{\prime}\,=\,-e^f \left\{\sqrt{2}e^{\varphi}e^{-2g}\frac{1}{\bar{g}}\,+\,\frac{\bar{g}}{6\sqrt{2}}\left(2e^{-\varphi}\,+\,e^{2\varphi}\right)\right\}, \\ \nonumber
 & & \varphi^{\prime}\,=\,-e^f \left\{\sqrt{2}e^{\varphi}e^{-2g}\frac{1}{\bar{g}}\,+\,\frac{\bar{g}}{3\sqrt{2}}\left(e^{-\varphi}\,-\,e^{2\varphi}\right)\right\}.
\end{eqnarray}
The boundary conditions as $r \rightarrow 0$ 
are 
$g(r),f(r) \rightarrow - Log(r)$ so that from (\ref{eq:sol3}) we get
\begin{equation}
\varphi(r) \rightarrow -\frac{1}{2} r^2 Log(r) + C_{\varphi}r^2 \ .
\label{cp}
\end{equation}
As in
 \cite{malda} there is a relation between $g$ and $\varphi$ which reads
\begin{equation}
  \label{eq:gfi}
  e^{2g+\varphi}\,=\,e^{2(g-\varphi)}\,+\,\frac{12}{\bar{g}^2}\left(g\,+\,\varphi\right)\,+\,C.
\end{equation}
The integration constant $C$ is related to the integration constant $C_{\varphi}$ (\ref{cp})
of the differential equation for $\varphi$, i.e. to the expectation value in the UV of the operator dual to $\varphi$ 
\cite{malda}. 
For an asymptotic configuration (\ref{eq:csol1}) we find $C=\frac{1}{4}\left(3-\log{4}\right)$.

\begin{figure}[ht]
\begin{center}
\epsfxsize=4in\leavevmode\epsfbox{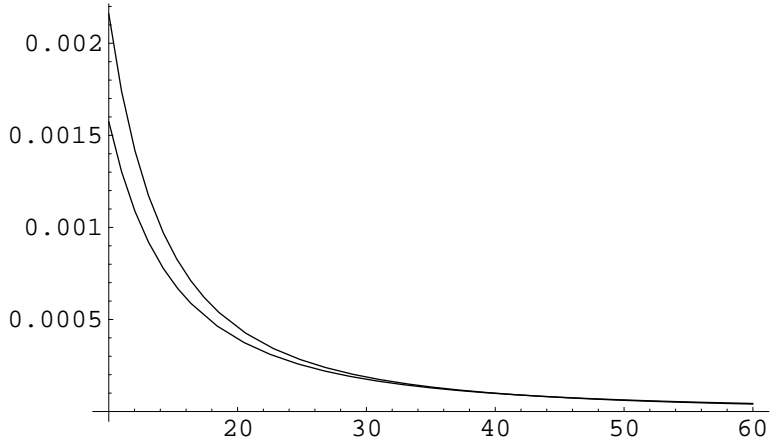}
\end{center}
\caption{In this figure we plot $e^{2 f(r)}$ as obtained from a numerical solution 
of equations (\ref{eq:sol3}) when $C=\frac{1}{4}\left(3-\log{4}\right)$ in comparison to the expected 
behaviour in the IR, 
$e^{2f}=\frac{1}{4^{\frac{4}{3}}r^2}$ in equation  (\ref{eq:csol1}).
We see that the two coincide in the IR.}
\label{phasefig1}
\end{figure}

We therefore expect to have for this value of $C$
a solution interpolating between  $AdS_5$ in the UV with $AdS_2 \times H^3$ in the IR.
Indeed we can find such a solution numerically.
In figure (\ref{phasefig1}) we plot $e^{2 f(r)}$ as obtained from a numerical solution 
of equations (\ref{eq:sol3}) when $C=\frac{1}{4}\left(3-\log{4}\right)$ in comparison to the expected 
behaviour in the IR, 
$e^{2f}=\frac{1}{4^{\frac{4}{3}}r^2}$ (\ref{eq:csol1}).
We see that indeed they coincide in the IR.

\begin{figure}[p]
\begin{center}
\epsfxsize=4in\leavevmode\epsfbox{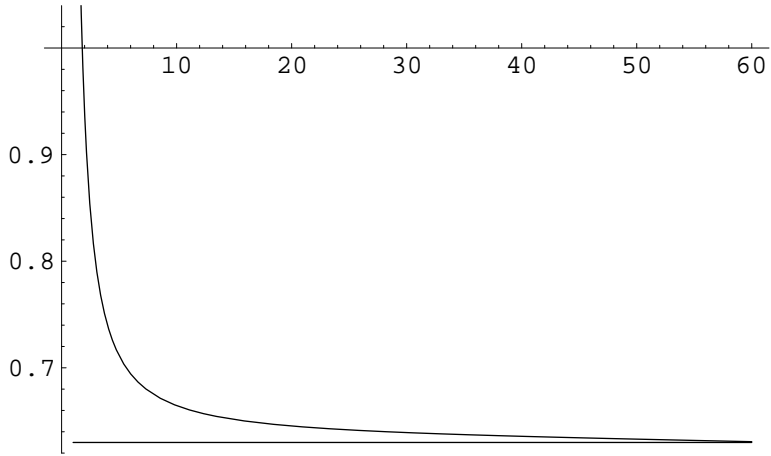}
\end{center}
\caption{In this figure  we plot $e^{2 g(r)}$ as obtained from a numerical solution 
in comparison to the expected 
behaviour in the IR, 
$e^{2g}=4^{-\frac{1}{3}}$  in equation  (\ref{eq:csol1}). We see 
that the two  coincide in the IR.}
\label{phasefig2}
\end{figure}

In figure (\ref{phasefig2}) we plot $e^{2 g(r)}$ as obtained from a numerical solution 
in comparison to the expected 
behaviour in the IR, 
$e^{2g}=4^{-\frac{1}{3}}$ (\ref{eq:csol1}). Again we see 
that they coincide in the IR.

 An expansion of (\ref{eq:gfi}) leads to $C_{\varphi}=\frac{C}{3}$ so that 
\begin{equation}
C_{\varphi}^{crit.}=\frac{1}{4}-\frac{\log{4}}{12} \ .
\label{cpc}
\end{equation}
As noted, we can interpret $C_{\varphi}$ in (\ref{cp})  roughly as  
the expectation value of 
the operator dual to $\varphi$.
$C_{\varphi}= C_{\varphi}^{crit.}$ is then
the value for which the Higgs and Coulomb branches ``intersect'' and this is where
we expect a fixed point.
For 
$C_{\varphi}> C_{\varphi}^{crit.}$ we move into the Coulomb branch while
for $C_{\varphi}< C_{\varphi}^{crit.}$ we move into the Higgs branch.
It is then of interest to see what happens to the supergravity solution as we
vary $C_{\varphi}$, which enters as a
 boundary value for $\varphi$. 
This analysis can be done numerically.

\begin{figure}[hp]
\begin{center}
\epsfxsize=4in\leavevmode\epsfbox{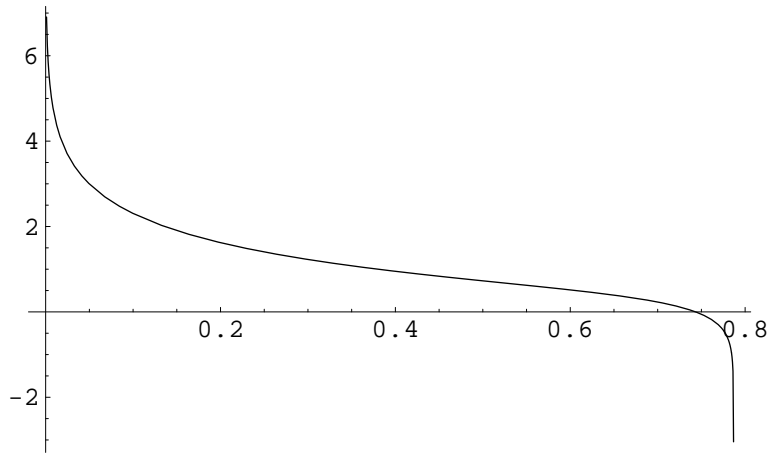}
\end{center}
\caption{The numerical 
behaviour of $g(r)$ approaching a ``bad'' singularity in the Coulomb branch,
$C_{\varphi}>C_{\varphi}^{crit.}$}
\label{phasefig3}
\end{figure}

\begin{figure}[hp]
\begin{center}
\epsfxsize=4in\leavevmode\epsfbox{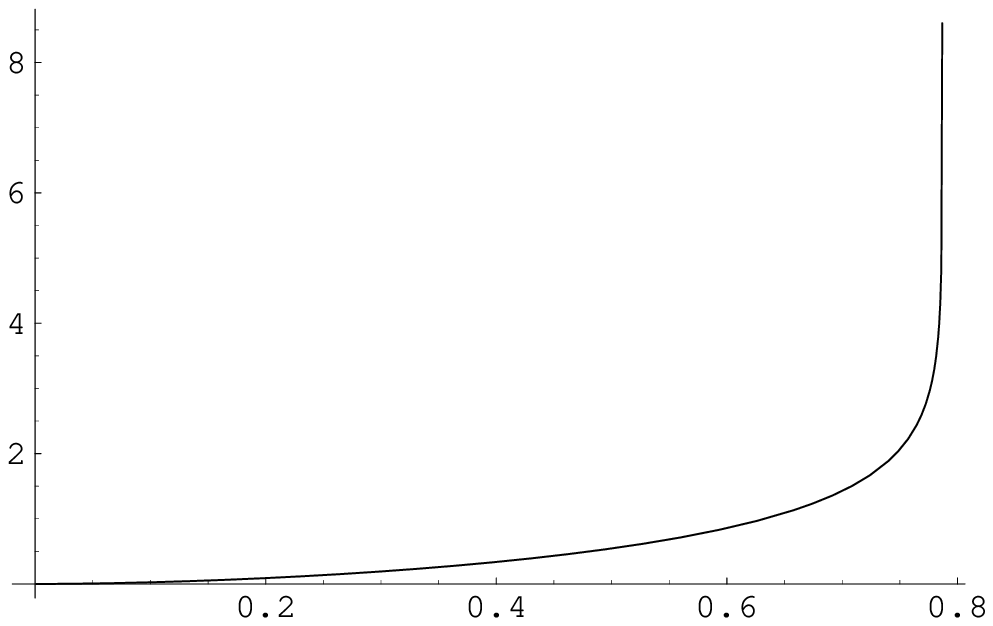}
\end{center}
\caption{$\varphi(r)$ blows up near a ``bad'' singularity in the Coulomb branch,
$C_{\varphi}>C_{\varphi}^{crit.}$}
\label{phasefig4}
\end{figure}

The numerical analysis of (\ref{eq:sol3}) shows singular solutions 
for $C_{\varphi}>C_{\varphi}^{crit.}$, as depicted in figure (\ref{phasefig3}) 
and  (\ref{phasefig4}).

We can check the nature of the singularity
according to the criterion of \cite{gubser}.
We need to compute the effective potential of the five-dimensional
 gauged supergravity corresponding  
to the above solution. For our ansatz with the gauge field determined by the twist we find from the Lagrangian given in \cite{romans}
\begin{equation}
  \label{eq:ep1}
  V_{eff}\,=\,\frac{3}{16}e^{2\varphi}e^{-4g}-\left(e^{-2\varphi}+2e^{\varphi}\right).
\end{equation}
Since this potential diverges near the singularity it is  of ``bad'' type.
The meaning of this singularity is that in the limit that we consider the Higgs and the Coulomb branches
decouple, and we remain at low-energy with a $(0+1)$-dimensional $\sigma$-model on the Higgs branch.

\begin{figure}[hp]
\begin{center}
\epsfxsize=3.5in\leavevmode\epsfbox{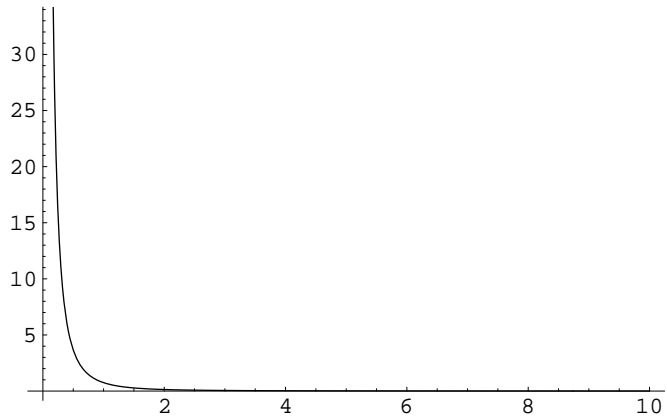}
\end{center}
\caption{The radius $e^{2f(r)}$ is shrinking in the IR, in
the Higgs branch 
$C_{\varphi}<C_{\varphi}^{crit.}$.
However the effective potential is bounded from above and the singularity 
is of a ``good'' type.} 
\label{phasefig5}
\end{figure}

\begin{figure}[hp]
\begin{center}
\epsfxsize=4in\leavevmode\epsfbox{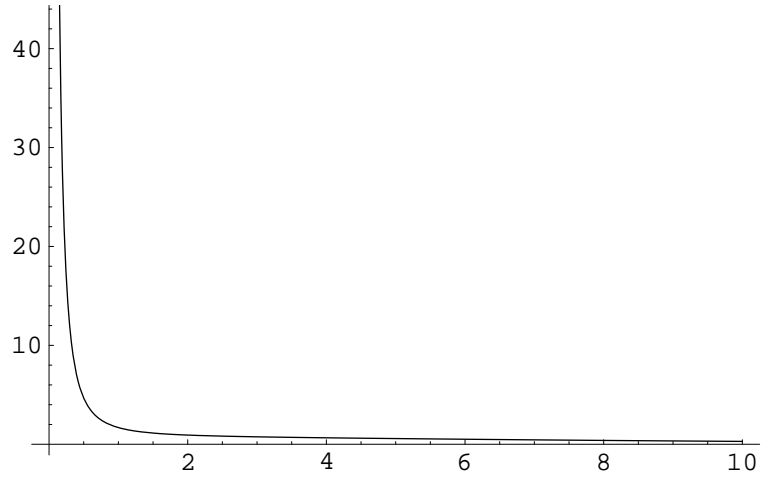}
\end{center}
\caption{The qualitative behaviour of the radius $e^{2g(r)}$ is very similar 
to that of $e^{2f(r)}$ in
the Higgs branch, 
$C_{\varphi}<C_{\varphi}^{crit.}$. It is also shrinking in the IR.}
\label{phasefig6}
\end{figure}

\begin{figure}[hp]
\begin{center}
\epsfxsize=4in\leavevmode\epsfbox{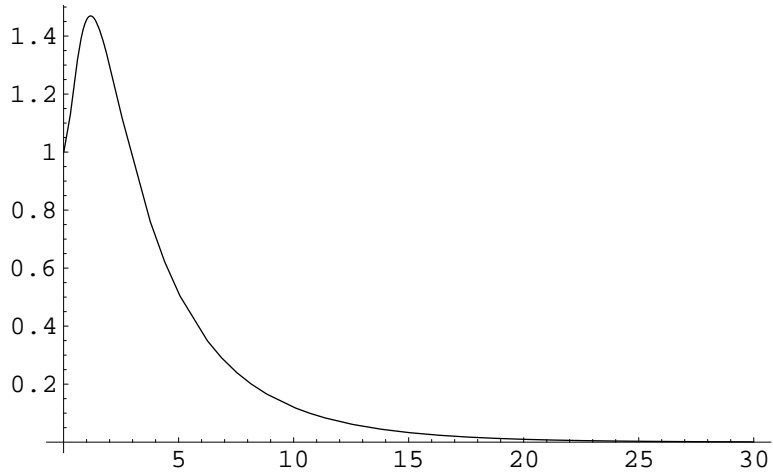}
\end{center}
\caption{ $e^{2\varphi(r)}$ in
the Higgs branch, $C_{\varphi}<C_{\varphi}^{crit.}$. 
It goes to zero as it approaches the ``good'' singularity in the IR. }
\label{phasefig7}
\end{figure}

The
numerical solution for $C_{\varphi}<C_{\varphi}^{crit.}$ has a different behaviour.
The radii $exp(2f(r))$ and $exp(2g(r))$ are shrinking in the IR 
as seen in figures (\ref{phasefig5}) and  (\ref{phasefig6}), which means increasing
of the curvature
of the solution.
However, in this case the behaviour of $\varphi$ is different as
seen in figure  (\ref{phasefig7})
and the singularity in the extreme IR is of a ``good'' type. 

\newpage

\subsection{$D3$-brane wrapped on $S^3$}
Consider now the $N$ $D3$-branes wrapped on an associative 3-cycle 
$S^3$.
Recall that for the five dimensional metric we take the ansatz
\begin{equation}
  \label{eq:metric1}
    ds^2=e^{2f}(-dt^2+dr^2) +e^{2g}(d\phi ^2 + 
\sin^2 \phi d \theta ^2 + \sin^2 \phi \sin^2\theta d \nu^2) \ ,
\end{equation}
where $f,g$ depend only on $r$.

The supergravity equations can be obtained from  (\ref{eq:sol3}).
In (\ref{eq:sol3}),
 due to the twist the first term is proportional to the scalar curvature of 
the supersymmetric cycle. The transition $H^3 \rightarrow S^3$ thus induces a minus sign in this term. This can also be seen via
 the coordinate transformation $\phi \rightarrow i \phi$ as argued in \cite{malda}. 
Hence we get the following equations for the $S^3$ case 
\begin{eqnarray}
  \label{eq:sol1}
  & & g^{\prime}\,=\,-e^f \left\{\sqrt{2}e^{\varphi}e^{-2g}\frac{1}{\bar{g}}\,+\, 
 \frac{\bar{g}}{6\sqrt{2}}\left(2e^{-\varphi}\,+\,e^{2\varphi}\right)\right\}, \\ \nonumber
  & & f^{\prime}\,=\,-e^f \left\{-\sqrt{2}e^{\varphi}e^{-2g}
\frac{1}{\bar{g}}\,+\, \frac{\bar{g}}{6\sqrt{2}}
\left(2e^{-\varphi}\,+\,e^{2\varphi}\right)\right\}, \\ \nonumber
 & & \varphi^{\prime}\,=\,-e^f \left\{-\sqrt{2}e^{\varphi}e^{-2g}
\frac{1}{\bar{g}}\,+\, \frac{\bar{g}}{3\sqrt{2}}\left(e^{-\varphi}\,-\,e^{2\varphi}\right)\right\}.
\end{eqnarray}

Again, the boundary conditions as $r \rightarrow 0$ 
are 
$g(r),f(r) \rightarrow - Log(r)$ so that from (\ref{eq:sol1}) we get
\begin{equation}
\varphi(r) \rightarrow \frac{1}{2} r^2 Log(r) + C_{\varphi}r^2 \ .
\label{cp1}
\end{equation}

\begin{figure}[hp]
\begin{center}
\epsfxsize=4in\leavevmode\epsfbox{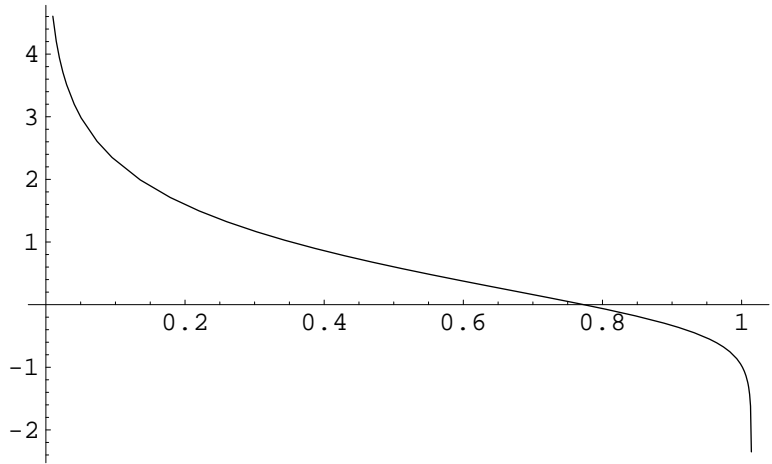}
\end{center}
\caption{A numerical solution for $g(r)$ in the case of $S^3$ cycle.}
\label{phasefig8}
\end{figure}

\begin{figure}[hp]
\begin{center}
\epsfxsize=4in\leavevmode\epsfbox{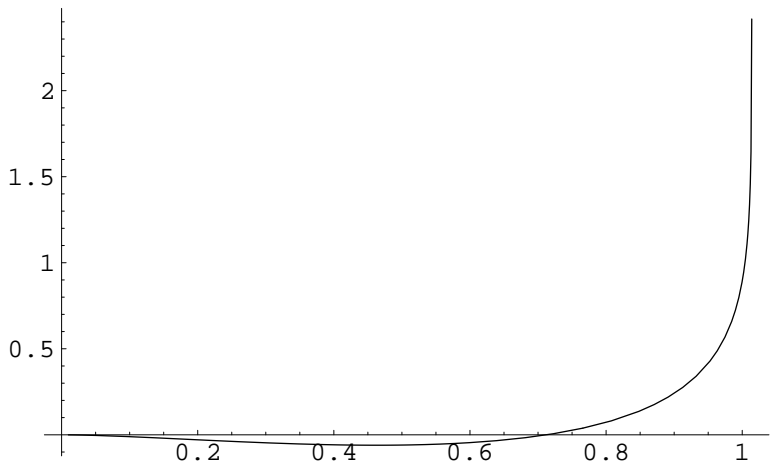}
\end{center}
\caption{A numerical solution of $\varphi(r)$ for $S^3$ cycle.}
\label{phasefig9}
\end{figure}

We can find a numerical solution as depicted in figures (\ref{phasefig8}) and
(\ref{phasefig9}). The numerical solution has the same behaviour independently of the value
of  $C_{\varphi}$. 
The 
 singular behaviour is as for the $H^3$ case on the Coulomb branch, i.e.
the singularity is of the ``bad'' type.

As before, the 
``bad'' singularity in the Coulomb branch
can be explained by the fact that in the limit that we consider the Higgs and the Coulomb branches
decouple, and we remain at low-energy with a $(0+1)$-dimensional $\sigma$-model on the Higgs branch.
However, 
in this case, unlike $H^3$, 
we do not expect a Higgs branch for 
the dual theory since there are no harmonic spinors on $S^3$ and
therefore it is rigid in the ambient seven-dimensional manifold.
This explains the difference between the type of singularity along the Higgs branch
in $H^3$ compared to $S^3$.

\newpage

\vskip 2cm  

{\bf Acknowledgement}:
We are grateful to  J. Maldacena for pointing out that the correct interpretation of
the supergravity solutions in section 4 is that of D3-branes wrapping associative 3-cycles and not 
special Lagrangian 3-cycles as written in the previous version of this paper.
The  $AdS_2 \times ds^2_{H^3}$ solution in section 4 has been independently obtained by J. Maldacena and
C. Nunez. We would like to thank J. Maldacena for this information and for other valuable discussions.

\newpage

\appendix{The Riemann surface case}

For the case of a 
$D3$-brane 
wrapped around a Riemann surface 
we get from (\ref{eq:vargravi}) and (\ref{eq:varsp2}) the solution of \cite{malda}. 
The metric ansatz reads
\begin{equation}
  \label{eq:maldametric}
  ds^2 \, =\, e^{2f}(dr^2\,+\,dz^2\,-\,dt^2)\,+\,\frac{e^{2g}}{y^2}(dx^2\,+\,dy^2)
\end{equation}
and the corresponding spin connection
\begin{eqnarray}
  \label{eq:spico}
  & & \omega^{34}_{x}\,=\,-\frac{1}{y}, \\ \nonumber
 & & \omega^{31}_{x}\,=\,\omega^{41}_{y}\,=\,\frac{e^{g-f}g^{\prime}}{y}, \\ \nonumber
 & & \omega^{21}_{z}\,=\,\omega^{01}_{t}\,=\,f^{\prime}, 
\end{eqnarray}
where the flat indices $(1,2,..)$ are connected with the curved ones by the vielbein
\begin{eqnarray}
  \label{eq:vielb}
  & & e^0_t\,=\,e^1_r\,=\,e^2_z\,=\,e^f, \\ \nonumber
  & & e^3_x\,=\,e^4_y\,=\,\frac{e^g}{y}.
\end{eqnarray}
We set $B_{\mu \nu}^{\alpha}$ to zero and take $\Phi$ to depend only on $r$, furthermore we truncate the  $SU(2) \times U(1)$ gauge group to the Cartan subgroup $U(1)\times U(1)$. From the spin connection (\ref{eq:spico}) we get the gauge potentials
\begin{equation}
  \label{eq:gfs}
  a_x\,=\,\frac{a}{y}\;,\; A_x^3\,=\,\frac{b}{y}
\end{equation}
and the corresponding field strengths
\begin{equation}
  \label{eq:fs}
  f_{xy}\,=\,\frac{a}{y^2}\;,\;F^3_{xy}\,=\,\frac{b}{y^2}.
\end{equation}
In addition we impose the following conditions on the spinors
\begin{equation}
  \label{eq:susyb}
  \gamma_{34}\, \varepsilon_a\,=\,-i\varepsilon_a\;,\;\gamma_{1}\,\varepsilon_a\,=\,\varepsilon_a,
\end{equation}
as in \cite{malda}. We rescale the coupling constants \cite{romans}, $g_1\,=\,\frac{\bar{g}}{\sqrt{2}}$ and $g_2\,=\,\bar{g}$. In this background with the above definitions the susy-variation of the $x$-component of the gravitino reads
\begin{eqnarray}
  \label{eq:gravi2}
  & & \delta\, \psi_{xa}\,=\,\partial_x \varepsilon_a\,+\,\frac{1}{2}\omega_x^{34}\gamma_{34}\,\varepsilon_a\,+\,\frac{1}{2}\omega_x^{31}\gamma_{31}\,\varepsilon_a\,+\,\frac{1}{2}\frac{\bar{g}}{2} \frac{a}{y}\left(\Gamma_{45}\right)_a^{\,b}\;\varepsilon_b\,+ \\ \nonumber 
 & & +\,\frac{1}{2}\bar{g}\frac{b}{y}\left(\Gamma_{345}\right)_a^{\,b}\;\varepsilon_b\,-\,i\gamma_x\,\frac{1}{12\sqrt{2}}\bar{g}\left(2e^{-\varphi}\,+\,e^{2\varphi}\right)\left(\Gamma_{45}\right)_a^{\,b}\;\varepsilon_b\,+ \\ \nonumber
 & & +\,i\frac{2}{3}\sqrt{\frac{1}{2}}\,\gamma^y\,e^{\varphi}\,F^3_{xy}\left(\Gamma_{3}\right)_a^{\,b}\;\varepsilon_b\,+\,i\frac{1}{3}\,e^{-2\varphi}\,f_{xy}\,\gamma^y\,\varepsilon_a,
\end{eqnarray}
where we have set $\varphi\, \equiv \, \sqrt{\frac{2}{3}}\phi$. Setting (\ref{eq:gravi2}) to zero yields
\begin{equation}
  \label{eq:constraints2}
  \bar{g} \left(\frac{a}{\sqrt{2}}\Gamma_{45}\,+\,b\Gamma_{345}\right)_a^{\;b}\varepsilon_b\,=\,-i\varepsilon_a
\end{equation}
and 
\begin{eqnarray}
  \label{eq:juppi}
  g^{\prime}\varepsilon_a\,=\,e^f\left(\frac{i}{6\sqrt{2}}\bar{g}\left(2e^{-\varphi}\,+\,e^{2\varphi}\right)\left(\Gamma_{45}\right)_a^{\,b}\;\varepsilon_b\,-\right. \\ \nonumber
 \left.-\,\frac{4}{3}\sqrt{\frac{1}{2}}\,e^{-2g}\,e^{\varphi}\,b\left(\Gamma_3\right)_a^{\,b}\;\varepsilon_b\,-\,\frac{2}{3}\,e^{-2g}\,e^{-2\varphi}\,a\,\varepsilon_a \right).
\end{eqnarray}
Eqn. (\ref{eq:constraints2}) can be solved in the following ways
\begin{equation}
  \label{eq:way1}
  \left(\Gamma_{45}\right)_a^{\,b}\;\varepsilon_b\,=\,i\varepsilon_a\, ,\,\left(\Gamma_3\right)_a^{\,b}\;\varepsilon_b\,=\,\pm \varepsilon_a\,\rightarrow\, \bar{g}\left(\frac{a}{\sqrt{2}}\,\pm\,b\right)\,=\,-1,
\end{equation}
\begin{equation}
  \label{eq:way3}
  \left(\Gamma_{45}\right)_a^{\,b}\;\varepsilon_b\,=\,-i\varepsilon_a\, ,\,\left(\Gamma_3\right)_a^{\,b}\;\varepsilon_b\,=\,\pm \varepsilon_a\,\rightarrow\, \bar{g}\left(\frac{a}{\sqrt{2}}\,\pm\,b\right)\,=\,1,
\end{equation}
Solutions which differ in the eigenvalue of $\Gamma_3$ are equivalent whereas solutions with different eigenvalues of $\Gamma_{45}$ give different solutions. Using (\ref{eq:way1}) we can solve (\ref{eq:juppi}) by choosing $b=0\,,\,a=-\frac{\sqrt{2}}{\bar{g}}$ to give
\begin{equation}
  \label{eq:sacrahaxn}
  g^{\prime}\,=\,e^f\left(-\frac{\bar{g}}{6\sqrt{2}}\left(2e^{-\varphi}\,+\,e^{2\varphi}\right)\,+\,\frac{2}{3}\,\frac{\sqrt{2}}{\bar{g}}\,e^{-2g}\,e^{-2\varphi}\right),
\end{equation}
which is part of the solution for the twist preserving $(4,4)$ supersymmetry in \cite{malda}. 
Exact correspondence is obtained if we set $\bar{g}\,=\,2\sqrt{2}$.

If we set $a=0\,,\, b=-\frac{1}{\bar{g}}$ we get the corresponding expression for the twist preserving $(2,2)$ supersymmetry
\begin{equation}
  \label{eq:alsdann}
  g^{\prime}\,=\,e^f\left(-\frac{1}{3}\left(2e^{-\varphi}\,+\,e^{2\varphi}\right)\,+\,\frac{1}{3}\,e^{-2g}\,e^{\varphi}\right).
\end{equation}
The solutions corresponding to (\ref{eq:way3}) give an overall sign in (\ref{eq:sacrahaxn}) and 
(\ref{eq:alsdann}) and will
 be  ruled out by the boundary conditions for $f,g$ in the limit $r \rightarrow 0$. 
By solving the remaining fermionic supersymmetry variations we get the differential equations for $f$ and $\varphi$.

The case $H^2$ resembles $H^3$ while 
 the $S^2$ case resembles the $S^3$ one.
In the $H^2$ case the equations read \cite{malda}
\begin{eqnarray}
  \label{eq:sincy}
  & & g^{\prime}\,=\,e^f\left(-\frac{1}{3}\left(2e^{-\varphi}\,+\,e^{2\varphi}\right)\,+\,\frac{1}{3}\,e^{-2g}\,e^{\varphi}\right), \\ \nonumber
  & & f^{\prime}\,=\,-\frac{e^f}{6}\left(2\left(2e^{-\varphi}\,+\,e^{2\varphi}\right)\,+\,e^{-2g}\,e^{\varphi}\right), \\ \nonumber
  & & \varphi^{\prime}\,=\,\frac{2}{3}e^f\left(\left(-e^{-\varphi}\,+\,e^{2\varphi}\right)\,-\,\frac{1}{4}\,e^{-2g}\,e^{\varphi}\right).
\end{eqnarray}

The boundary conditions as $r \rightarrow 0$ 
are 
$g(r),f(r) \rightarrow - Log(r)$ so that from (\ref{eq:sincy}) we get
\begin{equation}
\varphi(r) \rightarrow -\frac{1}{6} r^2 Log(r) + C_{\varphi}r^2 \ .
\label{cp2}
\end{equation}
There is a relation between $g$ and $\varphi$ which reads
\begin{equation}
  \label{eq:gfi1}
  e^{2g+\varphi}\,=\,e^{2(g-\varphi)}\,+\,\frac{1}{2}\left(g\,+\,2\varphi\right)\,+\,C.
\end{equation}
The $AdS_3 \times H^2$ solution of (\ref{eq:sincy}) gives $C=\frac{1}{4}$.

We therefore expect to have for this value of $C$
a solution interpolating between  $AdS_5$ in the UV with $AdS_3 \times H^2$ in the IR and indeed we find one. Similar solutions were constructed in \cite{sabra1} in the context of $D=5, N=2$ gauged supergravity which were generalized in \cite{sabra2} where also an analysis of solutions of the form $M_3 \times M_2$ was given.

 An expansion of (\ref{eq:gfi1}) leads to $C_{\varphi}=\frac{C}{3}$ so that 
\begin{equation}
C_{\varphi}^{crit.}=\frac{1}{12} \ .
\label{cpc1}
\end{equation}
Again, one interprets $C_{\varphi}$ in (\ref{cp2})  roughly as  
the expectation value of 
the operator dual to $\varphi$.
$C_{\varphi}= C_{\varphi}^{crit.}$ is then
the value for which the Higgs and Coulomb branches ``intersect'' and this  is  where
we expect a fixed point.
For 
$C_{\varphi}> C_{\varphi}^{crit.}$ we move into the Coulomb branch while
for $C_{\varphi}< C_{\varphi}^{crit.}$ we move into the Higgs branch.
It is then of interest to see what happens to the supergravity solution as we
vary $C_{\varphi}$, which enters as a
 boundary value for $\varphi$. 
The numerical analysis of (\ref{eq:sincy}) shows singular solutions
in both cases.
However, the nature of the singularity is different and like the $H^3$ case it is of
the ``good'' type along the Higgs
 branch and of the ``bad'' type along the Coulomb branch, i.e.
the potential
\begin{equation}
  \label{eq:poteff}
  V_{eff}\,=\,\frac{1}{4}e^{2\varphi}e^{-4g}-4\left(e^{-2\varphi}+2e^{\varphi}\right).
\end{equation}
is unbounded from above near the singularity.

In the case of $S^2$ we find, like in the $S^3$ case,
 that the singularity is of the ``bad'' type independently of the
value of  $C_{\varphi}$.

As before, the 
``bad'' singularity in the Coulomb branch
can be explained by the fact that in the limit that we consider the Higgs and the Coulomb branches
decouple, and we remain at low-energy with a $(1+1)$-dimensional $\sigma$-model on the Higgs branch.
The ``bad'' singularity on the Higgs branch in the 
$S^2$ case, indicates its absence.

\appendix{The $S^3,H^3$ examples}

For the five dimensional metric we take the ansatz
\begin{equation}
\label{eq:metric2}
    ds^2=e^{2f}(-dt^2+dr^2) +e^{2g}(d\phi ^2 + \sin^2 \phi d \theta ^2 + \sin^2 \phi \sin^2\theta d \nu^2),
\end{equation}
where $f,g$  depend only on $r$. The second part of the above metric is the metric of the $S^3$.

The components of the vielbein of this metric read
\begin{eqnarray}
  \label{eq:mv1}
  & & e_t^0=e^f, \\  \nonumber
  & & e_r^1=e^f, \\  \nonumber
  & & e_{\phi}^2=e^g , \\  \nonumber
  & & e_{\theta}^3=e^g  \sin \phi, \\  \nonumber
  & & e_{\nu}^4=e^g  \sin \phi \sin \theta.
\end{eqnarray}

For the boundary conditions to be imposed on the $SU(2)$ gauge potentials we need the spin connection on the $S^3$. For the $S^3$ part of the above metric we get the following nonvanishing connection coefficients
\begin{eqnarray}
  \label{eq:christoffels}
  & & \Gamma_{\theta \theta}^{\phi}= - \sin \phi \cos \phi,  \\ \nonumber
  & & \Gamma_{\theta \phi}^{\theta}= \cot\phi, \\ \nonumber
  & & \Gamma_{\nu \nu}^{\phi}= - \sin^2\theta \sin \phi \cos \phi, \\ \nonumber
  & & \Gamma_{\nu \phi}^{\nu}= \cot\phi, \\ \nonumber
  & & \Gamma_{\nu \nu}^{\theta}= - \sin \theta \cos \theta,  \\ \nonumber
  & & \Gamma_{\nu \theta}^{\nu}= \cot\theta.
\end{eqnarray}
With the vielbeins
\begin{eqnarray}
  \label{eq:vielbein}
  & & e_{\phi}^{\bar{1}}=1, \\ \nonumber
  & & e_{\theta}^{\bar{2}}=\sin \phi, \\ \nonumber
  & & e_{\nu}^{\bar{3}}=\sin \phi \sin \theta, 
\end{eqnarray}
we get for the spin connection 
\begin{eqnarray}
  \label{eq:spinc}
  & & \omega_{\theta}^{\bar{1}\bar{2}}=-\cos \phi = \omega_{\theta}^{23}, \\ \nonumber 
  & & \omega_{\nu}^{\bar{1}\bar{3}}=-\sin \theta \cos \phi =\omega_{\nu}^{24}, \\ \nonumber
  & & \omega_{\nu}^{\bar{2}\bar{3}}=-\cos \theta =\omega_{\nu}^{34},
\end{eqnarray}
where the barred superscripts denote flat indices on $S^3$.
Thus we are led to the gauge potentials
\begin{eqnarray}
  \label{eq:gp1}
  & & A_{\theta}^{1}=a\cos \phi, \\ \nonumber
  & & A_{\nu}^{2}=b\cos \theta, \\ \nonumber
  & & A_{\nu}^{3}= c\sin \theta \cos \phi,
\end{eqnarray}
(identifying $23\, \leftrightarrow \, 1, \, 34 \, \leftrightarrow \, 2, \, 24 \, \leftrightarrow \,3$).
The only nonvanishing components of the corresponding field strength are
\begin{eqnarray}
  \label{eq:gf1}
  & & F^{1}_{\theta \phi}=a\sin \phi, \\ \nonumber
  & & F^{2}_{\nu \theta}=\sin \theta (b+\bar{g}ac\cos^2 \phi), \\ \nonumber
  & & F^{3}_{\nu \phi}=c\sin \theta \sin \phi, \\ \nonumber
  & & F^{3}_{\nu \theta}=-\cos \theta \cos \phi(c+\bar{g}ab), 
\end{eqnarray}
where the superscripts refer to the generators $T_a$ of $SU(2)$, which is suitably embedded in the $Spin(5)$ $R$-symmetry group, with the Lie algebra $[T_a,T_b]=\epsilon_{abc} T_c$. The coupling constant $\bar{g}$ is included for later reference. The other nonvanishing connection coefficients for the metric (\ref{eq:metric2}) are
\begin{eqnarray}
  \label{eq:christoffels2}
  & & \Gamma_{t t}^{r}= f^{\prime}, \\ \nonumber
  & & \Gamma_{t r}^{t}=f^{\prime}, \\ \nonumber
  & & \Gamma_{\phi \phi}^r= -e^{2(g-f)}g^{\prime}, \\ \nonumber
  & & \Gamma_{\theta \theta}^r= - e^{2(g-f)}g^{\prime}\sin^2\phi, \\ \nonumber
  & & \Gamma_{\nu \nu}^r= -e^{2(g-f)}g^{\prime} \sin^2 \phi \sin^2 \theta, \\ \nonumber
  & & \Gamma_{\phi r}^{\phi}=\Gamma_{\theta r}^{\theta}=\Gamma_{\nu r}^{\nu}=g^{\prime}, \\ \nonumber
  & & \Gamma_{r r}^r=f^{\prime},
\end{eqnarray}
while calculating the spin connection yields
\begin{eqnarray}
  \label{eq:spinc2}
  & & \omega_t^{10}=-f^{\prime}, \\ \nonumber
  & & \omega_{\phi}^{12}=-e^{g-f}g^{\prime}, \\ \nonumber
  & & \omega_{\theta}^{13}=-e^{g-f}g^{\prime}\sin \phi, \\ \nonumber
  & & \omega_{\nu}^{14}=-e^{g-f}g^{\prime}\sin \phi \sin \theta.
\end{eqnarray}

The variation of the fermionic fields is given by

\begin{eqnarray}
  \label{eq:expl1}
  & & \delta \psi_{t a}\,= \\ \nonumber
 & & =\, \partial_{t}\, \varepsilon_a\,+\,\frac{1}{2}f^{\prime}\gamma_0\, 
\varepsilon_a\,-\,i e^f\, \gamma_0\,\frac{1}{12\sqrt{2}}\bar{g}
\left(2e^{-\varphi}+e^{2\varphi}\right) 
\left(\Gamma_{45}\right)_a^{\;b}\varepsilon_b\;+ \\ \nonumber
 & & +\,i\;\frac{1}{3}\;\sqrt{\, \frac{1}{2}
}e^{\varphi}e^{(f-2g)}\,\sin^{-1} \phi 
\left\{\gamma_{023}\;F^1_{\theta \phi}\left(\Gamma_1\right)_a^{\;b}\varepsilon_b\,+\right. \\ \nonumber
& & \left.+\sin^{-1} \phi \sin^{-1} \theta 
\gamma_{034}\left(F^2_{\nu \theta}
\left(\Gamma_2\right)_a^{\;b}\varepsilon_b\,
+F^3_{\nu \theta}\left(\Gamma_3\right)_a^{\;b}\varepsilon_b\right)\,
+\,\sin^{-1} \theta \gamma_{024}\;F^3_{\nu \phi}\left(\Gamma_3\right)_a^{\;b}\varepsilon_b\right\}\,=\,0,
\end{eqnarray}
\begin{eqnarray}
  \label{eq:explphi}
   & & \delta \psi_{\phi a}\,= \\ \nonumber
& &  =\,\partial_{\phi}\, \varepsilon_a\,+\,\frac{1}{2}e^{g-f}
g^{\prime}\gamma_2\, \varepsilon_a\,-\,i e^g
\, \gamma_2\,\frac{1}{12\sqrt{2}}\bar{g}\left(2e^{-\varphi}+e^{2\varphi}\right) 
\left(\Gamma_{45}\right)_a^{\;b}\varepsilon_b\;+ \\ \nonumber
 & & +\,i\;\frac{1}{3}\;\sqrt{\, \frac{1}{2}}e^{\varphi}e^{-g}
\,\sin^{-2} \phi \sin^{-1} \theta 
\gamma_{234}\left(\;F^2_{\nu \theta}
\left(\Gamma_2\right)_a^{\;b}\varepsilon_b\,+
\,F^3_{\nu \theta}\left(\Gamma_3\right)_a^{\;b}\varepsilon_b\right)\,+ \\ \nonumber
& & + \,i\;\frac{2}{3}\;\sqrt{\, \frac{1}{2}}
e^{\varphi}e^{-g}
\,\sin^{-1} \phi\left( \sin^{-1} \theta \gamma_{4} 
\,F^3_{\phi \nu}\left(\Gamma_3\right)_a^{\;b}\varepsilon_b\,
+\,\gamma_3 \,F^1_{\phi \theta}\left(\Gamma_1\right)_a^{\;b}\varepsilon_b\right)\,=\,0,
\end{eqnarray}
\begin{eqnarray}
  \label{eq:expltheta}
   & & \delta \psi_{\theta a}\,= \\ \nonumber
& &  =\,\partial_{\theta}\, \varepsilon_a\,-\,
\frac{1}{2}\cos \phi \gamma_{23}\, \varepsilon_a\,+
\,\frac{1}{2}e^{g-f}g^{\prime}\sin \phi \gamma_3\, 
\varepsilon_a\,+\,\frac{1}{2}\bar{g} a \cos \phi 
\left(\Gamma_{145}\right)_a^{\;b}\,\varepsilon_b\,- \\ \nonumber
& & -\,i e^g \sin \phi \, \gamma_3\,\frac{1}{12\sqrt{2}}
\bar{g}\left(2e^{-\varphi}+e^{2\varphi}\right) 
\left(\Gamma_{45}\right)_a^{\;b}\varepsilon_b\,-\,i\;\frac{1}{3}\;\sqrt{\, 
\frac{1}{2}}e^{\varphi}e^{-g}\, \sin^{-1} \theta 
\gamma_{234}\;F^3_{\nu \phi}\left(\Gamma_3\right)_a^{\;b}\varepsilon_b\,+ 
\\ \nonumber
& & + \,i\;\frac{2}{3}\;\sqrt{\, \frac{1}{2}}e^{\varphi}e^{-g}\, 
\left\{ \gamma_2 \,F^1_{\theta \phi}\left(\Gamma_1\right)_a^{\;b}
\varepsilon_b\,+\sin^{-1} \phi \sin^{-1} 
\theta\,\gamma_4 \left(F^2_{\theta \nu}\left(\Gamma_2\right)_a^{\;b}
\varepsilon_b\,+\,F^3_{\theta \nu}\left(\Gamma_3\right)_a^{\;b}\varepsilon_b\right)\right\}\,=\,0,
\end{eqnarray}
\begin{eqnarray}
  \label{eq:explnu}
   & & \delta \psi_{\nu a}\,= \\ \nonumber
& &  =\,\partial_{\nu}\, \varepsilon_a\,-\,\frac{1}{2}\sin 
\theta \cos \phi \gamma_{24}\, \varepsilon_a\,-\,\frac{1}{2}\cos 
\theta \gamma_{34}\, \varepsilon_a\,+\,\frac{1}{2}e^{g-f}
g^{\prime}\sin \phi \sin \theta \gamma_4\, \varepsilon_a\,+ \\ \nonumber
& & +\,\frac{1}{2}\bar{g} b \cos \theta \left(\Gamma_{245}\right)_a^{\;b}\,
\varepsilon_b\,+\,\frac{1}{2}\bar{g} c \sin \theta \cos \phi 
\left(\Gamma_{345}\right)_a^{\;b}\,\varepsilon_b\,- \\ \nonumber
& & -\,i e^g \sin \phi \sin \theta \, \gamma_4\,\frac{1}{12\sqrt{2}}
\bar{g}\left(2e^{-\varphi}+e^{2\varphi}\right) \left(\Gamma_{45}\right)_a^{\;b}\varepsilon_b\,+
\,i\;\frac{1}{3}\;\sqrt{\, \frac{1}{2}}e^{\varphi}e^{-g}
\, \sin \theta \gamma_{234}\;
F^1_{\theta \phi} \left(\Gamma_1\right)_a^{\;b}\varepsilon_b\,+ \\ \nonumber \
& & +\,i\;\frac{2}{3}\;\sqrt{\, \frac{1}{2}}e^{\varphi}e^{-g}\, 
\left\{ \sin^{-1} \phi \gamma_3 \left( F^2_{\nu \theta}
\left(\Gamma_2\right)_a^{\;b}\varepsilon_b\,+\,F^3_{\nu \theta}
\left(\Gamma_3\right)_a^{\;b}\varepsilon_b \right)\,+\,
\gamma_2 F^3_{\nu \phi}\left(\Gamma_3\right)_a^{\;b}\varepsilon_b \right\}\,=\,0.
\end{eqnarray}

The $H^3$ case is obtained from the above by $sin(\phi) \rightarrow sinh(\phi)$.

\end{document}